\documentclass[12pt]{article}

\usepackage{amsbsy,amsfonts,amsmath,amssymb,amsthm,enumerate}
\usepackage{graphicx}
\usepackage{bm}

\newcommand{\beq}{\begin{equation}}
\newcommand{\eeq}{\end{equation}}
\newcommand{\beqs}{\begin{eqnarray}}
\newcommand{\eeqs}{\end{eqnarray}}

\newtheorem{lemma}{Lemma}[section]
\newtheorem{defi}{Definition}[section]
\newtheorem{conj}{Conjecture}[section]
\newtheorem{propo}{Proposition}[section]

\begin{document}

\title{Ice model and eight-vertex model on the two-dimensional Sierpinski gasket}

\author{
Shu-Chiuan~Chang\footnote{Department of Physics, National Cheng Kung University, Tainan 70101, Taiwan. {\tt scchang@mail.ncku.edu.tw}}
\footnote{Physics Division, National Center for Theoretical Science,
National Taiwan University, Taipei 10617, Taiwan},
Lung-Chi~Chen\footnote{Department of Mathematics, Fu-Jen Catholic
University, Taipei 24205, Taiwan. {\tt lcchen@math.fju.edu.tw}}
                            and
Hsin-Yun Lee \footnote{Department of Mathematics, Fu-Jen Catholic
University, Taipei 24205, Taiwan.  {\tt jammie@yahoo.com.tw}}
                            }

\date{\today}

\maketitle

\begin{abstract}

We present the numbers of ice model and eight-vertex model configurations (with Boltzmann factors equal to one), $I(n)$ and $E(n)$ respectively, on the two-dimensional Sierpinski gasket $SG(n)$ at stage $n$. For the eight-vertex model, the number of configurations is $E(n)=2^{3(3^n+1)/2}$ and the entropy per site, defined as $\lim_{v \to \infty} \ln E(n)/v$ where $v$ is the number of vertices on $SG(n)$, is exactly equal to $\ln 2$. For the ice model, the upper and lower bounds for the entropy per site $\lim_{v \to \infty} \ln I(n)/v$ are derived in terms of the results at a certain stage. As the difference between these bounds converges quickly to zero as the calculated stage increases, the numerical value of the entropy can be evaluated with more than a hundred significant figures accurate. The corresponding result of ice model on the generalized two-dimensional Sierpinski gasket $SG_b(n)$ with $b=3$ is also obtained. For the generalized vertex model on $SG_3(n)$, the number of configurations is $2^{(8 \times 6^n +7)/5}$ and the entropy per site is equal to $\frac87 \ln 2$. The general upper and lower bounds for the entropy per site for arbitrary $b$ are conjectured. 


\end{abstract}


\section{Introduction}
\label{sectionI}

The ice model was introduced by Pauling to study the residual entropy of water ice \cite{pauling35}, and was solved exactly by Lieb on the square lattice \cite{lieb67,lieb67n}. The eight-vertex model is a generalization of the ice-type (six-vertex) models \cite{sutherland70, wu70} and solved by Baxter for the zero-field case \cite{baxter71,baxterbook}. On the triangular lattice, there are 20 vertex configurations for the ice rule \cite{baxter69} and the 32-vertex model was considered \cite{wu75}. There is a correspondence between such model and the Ising model, while there are other related models, see, for example, \cite{DG1}.
It is of interest to consider the ice model and eight-vertex model on self-similar fractal lattices which have scaling invariance rather than translational invariance. Fractals are geometric structures of non-integer Hausdorff dimension realized by repeated construction of an elementary shape on progressively smaller length scales \cite{mandelbrot, Falconer}. A well-known example of fractal is the Sierpinski gasket which has been extensively studied in several contexts \cite{Gefen80, Gefen81, Rammal, Alexander, Domany, Gefen8384, Guyer, Kusuoka, Dhar97, Daerden, Dhar05, sts, sfs, ds, dms, css, hs}. We shall derive the recursion relations for the numbers of ice model and eight-vertex model configurations with Boltzmann factors equal to one on the two-dimensional Sierpinski gasket, and determine the entropies. We shall also consider the number of ice model configurations on a generalized two-dimensional Sierpinski gasket.

\section{Preliminaries}
\label{sectionII}

We first recall some relevant definitions in this section. A connected graph (without loops) $G=(V,E)$ is defined by its vertex (site) and edge (bond) sets $V$ and $E$ \cite{bbook,fh}.  Let $v(G)=|V|$ be the number of vertices and $e(G)=|E|$ the number of edges in $G$.  The degree or coordination number $k_i$ of a vertex $v_i \in V$ is the number of edges attached to it.  A $k$-regular graph is a graph with the property that each of its vertices has the same degree $k$. Let us consider a 4-regular graph first, and assign an orientation on each edge. For the ice model, the ice rule must be satisfied at every vertex. Namely, the number of arrows pointing inward at each vertex must be two and the number of arrows pointing outward is also two. There are six possible different configurations of arrows at each vertex. For the eight-vertex model, the number of arrows pointing inward at each vertex must be an even number. There are now eight possible different configurations of arrows at each vertex, including the six configurations of the ice model plus sink (all four arrows pointing inward) and source (all four arrows pointing outward). For the vertex with degree six, the number of arrows pointing inward and the number of arrows pointing outward are both three. There are twenty different arrow configurations at such vertex. For the 32-vertex model, the number of arrows pointing inward at degree-6 vertex must be an odd number. The thirty-two different arrow configurations includes the twenty configurations of the ice model plus only one arrow (six possibility) pointing outward and only one arrow (six possibility) pointing inward. In general, one can associate an energy to the vertex for each configuration. All such Boltzmann weights are set to one throughout this paper. 

Let us denote the total number of ice model configurations on a graph $G$ as $I(G)$ and that of eight-vertex model configurations as $E(G)$. The entropy per site for the ice model is given by
\beq
S_{I,G} = \lim_{v(G) \to \infty} \frac{\ln I(G)}{v(G)} \ ,
\label{zdef}
\eeq
where $G$, when used as a subscript in this manner, implicitly refers to
the thermodynamic limit. Similarly, the corresponding entropy per site for the eight-vertex model is denoted as $S_{E,G}$. We will see that the limits $S_{I,G}$ and $S_{E,G}$ exist for the Sierpinski gasket considered in this paper.

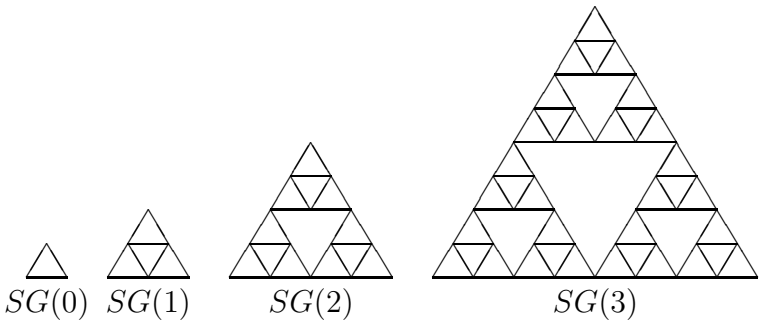
\begin{figure}[htbp]
\unitlength 0.9mm \hspace*{3mm}
\begin{picture}(108,40)
\put(0,0){\line(1,0){6}}
\put(0,0){\line(3,5){3}}
\put(6,0){\line(-3,5){3}}
\put(3,-4){\makebox(0,0){$SG(0)$}}
\put(12,0){\line(1,0){12}}
\put(12,0){\line(3,5){6}}
\put(24,0){\line(-3,5){6}}
\put(15,5){\line(1,0){6}}
\put(18,0){\line(3,5){3}}
\put(18,0){\line(-3,5){3}}
\put(18,-4){\makebox(0,0){$SG(1)$}}
\put(30,0){\line(1,0){24}}
\put(30,0){\line(3,5){12}}
\put(54,0){\line(-3,5){12}}
\put(36,10){\line(1,0){12}}
\put(42,0){\line(3,5){6}}
\put(42,0){\line(-3,5){6}}
\multiput(33,5)(12,0){2}{\line(1,0){6}}
\multiput(36,0)(12,0){2}{\line(3,5){3}}
\multiput(36,0)(12,0){2}{\line(-3,5){3}}
\put(39,15){\line(1,0){6}}
\put(42,10){\line(3,5){3}}
\put(42,10){\line(-3,5){3}}
\put(42,-4){\makebox(0,0){$SG(2)$}}
\put(60,0){\line(1,0){48}}
\put(72,20){\line(1,0){24}}
\put(60,0){\line(3,5){24}}
\put(84,0){\line(3,5){12}}
\put(84,0){\line(-3,5){12}}
\put(108,0){\line(-3,5){24}}
\put(66,10){\line(1,0){12}}
\put(90,10){\line(1,0){12}}
\put(78,30){\line(1,0){12}}
\put(72,0){\line(3,5){6}}
\put(96,0){\line(3,5){6}}
\put(84,20){\line(3,5){6}}
\put(72,0){\line(-3,5){6}}
\put(96,0){\line(-3,5){6}}
\put(84,20){\line(-3,5){6}}
\multiput(63,5)(12,0){4}{\line(1,0){6}}
\multiput(66,0)(12,0){4}{\line(3,5){3}}
\multiput(66,0)(12,0){4}{\line(-3,5){3}}
\multiput(69,15)(24,0){2}{\line(1,0){6}}
\multiput(72,10)(24,0){2}{\line(3,5){3}}
\multiput(72,10)(24,0){2}{\line(-3,5){3}}
\multiput(75,25)(12,0){2}{\line(1,0){6}}
\multiput(78,20)(12,0){2}{\line(3,5){3}}
\multiput(78,20)(12,0){2}{\line(-3,5){3}}
\put(81,35){\line(1,0){6}}
\put(84,30){\line(3,5){3}}
\put(84,30){\line(-3,5){3}}
\put(84,-4){\makebox(0,0){$SG(3)$}}
\end{picture}

\vspace*{5mm}
\caption{\footnotesize{The first four stages $n=0,1,2,3$ of the two-dimensional Sierpinski gasket $SG(n)$.}} 
\label{sgfig}
\end{figure}

The construction of the two-dimensional Sierpinski gasket $SG(n)$ at stage $n$ is shown in Fig. \ref{sgfig}. At stage $n=0$, it is an equilateral triangle; while stage $n+1$ is obtained by the juxtaposition of three $n$-stage structures. The two-dimensional Sierpinski gaskets has fractal dimensionality $D=\ln3/\ln2$ \cite{Gefen81}, and the numbers of edges and vertices are given by 
\beq
e(SG(n)) = 3^{n+1} \ ,
\label{e}
\eeq
\beq
v(SG(n)) = \frac{3}{2} [3^n+1] \ .
\label{v}
\eeq
Except the 3 outmost vertices which have degree 2, all other vertices of $SG(n)$ have degree 4. In the large $n$ limit, $SG$ is 4-regular. 

\bigskip

\bigskip

\begin{figure}[htbp]
\unitlength 0.9mm \hspace*{3mm}
\begin{picture}(108,45)
\put(0,0){\line(1,0){18}}
\put(3,5){\line(1,0){12}}
\put(6,10){\line(1,0){6}}
\put(0,0){\line(3,5){9}}
\put(6,0){\line(3,5){6}}
\put(12,0){\line(3,5){3}}
\put(18,0){\line(-3,5){9}}
\put(12,0){\line(-3,5){6}}
\put(6,0){\line(-3,5){3}}
\put(9,-4){\makebox(0,0){$SG_3(1)$}}
\put(24,0){\line(1,0){54}}
\put(33,15){\line(1,0){36}}
\put(42,30){\line(1,0){18}}
\put(24,0){\line(3,5){27}}
\put(42,0){\line(3,5){18}}
\put(60,0){\line(3,5){9}}
\put(78,0){\line(-3,5){27}}
\put(60,0){\line(-3,5){18}}
\put(42,0){\line(-3,5){9}}
\multiput(27,5)(18,0){3}{\line(1,0){12}}
\multiput(30,10)(18,0){3}{\line(1,0){6}}
\multiput(30,0)(18,0){3}{\line(3,5){6}}
\multiput(36,0)(18,0){3}{\line(3,5){3}}
\multiput(36,0)(18,0){3}{\line(-3,5){6}}
\multiput(30,0)(18,0){3}{\line(-3,5){3}}
\multiput(36,20)(18,0){2}{\line(1,0){12}}
\multiput(39,25)(18,0){2}{\line(1,0){6}}
\multiput(39,15)(18,0){2}{\line(3,5){6}}
\multiput(45,15)(18,0){2}{\line(3,5){3}}
\multiput(45,15)(18,0){2}{\line(-3,5){6}}
\multiput(39,15)(18,0){2}{\line(-3,5){3}}
\put(45,35){\line(1,0){12}}
\put(48,40){\line(1,0){6}}
\put(48,30){\line(3,5){6}}
\put(54,30){\line(3,5){3}}
\put(54,30){\line(-3,5){6}}
\put(48,30){\line(-3,5){3}}
\put(48,-4){\makebox(0,0){$SG_3(2)$}}
\put(84,0){\line(1,0){24}}
\put(87,5){\line(1,0){18}}
\put(90,10){\line(1,0){12}}
\put(93,15){\line(1,0){6}}
\put(84,0){\line(3,5){12}}
\put(90,0){\line(3,5){9}}
\put(96,0){\line(3,5){6}}
\put(102,0){\line(3,5){3}}
\put(108,0){\line(-3,5){12}}
\put(102,0){\line(-3,5){9}}
\put(96,0){\line(-3,5){6}}
\put(90,0){\line(-3,5){3}}
\put(96,-4){\makebox(0,0){$SG_4(1)$}}
\end{picture}

\vspace*{5mm}
\caption{\footnotesize{The generalized two-dimensional Sierpinski gasket $SG_b(n)$ with $b=3$ at stage $n=1, 2$ and $b=4$ at stage $n=1$.}} 
\label{sgbfig}
\end{figure}
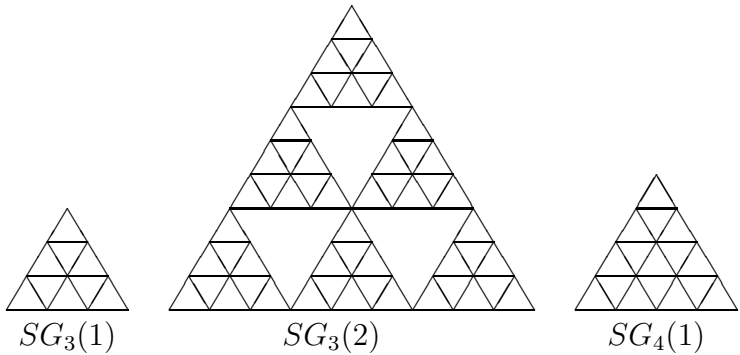

The two-dimensional Sierpinski gasket can be generalized, denoted as $SG_b(n)$, by introducing the side length $b$ which is an integer larger or equal to two \cite{Hilfer}. The generalized two-dimensional Sierpinski gasket at stage $n+1$ is constructed with $b$ layers of stage $n$ structures. The two-dimensional $SG_b(n)$ with $b=3$ at stage $n=1, 2$ and $b=4$ at stage $n=1$ are illustrated in Fig. \ref{sgbfig}. The ordinary two-dimensional Sierpinski gasket $SG(n)$ corresponds to the $b=2$ case, where the index $b$ is neglected for simplicity. The Hausdorff dimension for $SG_b$ is given by $D=\ln {b+1 \choose 2} / \ln b$ \cite{Hilfer}. Notice that $SG_b$ is not $k$-regular even in the thermodynamic limit. We shall use simplified notations $I_b(n)$ and $E_b(n)$ for the numbers of ice model and generalized vertex model configurations on $SG_b(n)$.

\bigskip

\bigskip

\section{The number of ice model configurations on $SG(n)$}
\label{sectionIII}

Denote the number of ice model configurations on the two-dimensional Sierpinski gasket $SG(n)$ as $I(n)$. In this section we derive its entropy per site in detail. As the three outmost vertices of the Sierpinski gasket have degree two, they are exempt from the ice rule in the calculation of $I(n)$. For the two edges connected to each of these outmost vertices, each of them can be directed either inward or outward independently. Let us define the quantities to be used.

\bigskip

\begin{figure}[htbp]
\unitlength 3mm 
\begin{picture}(44,3)
\put(0,0){\line(1,0){4}}
\put(0,0){\line(2,3){2}}
\put(4,0){\line(-2,3){2}}
\put(0,0){\vector(1,0){1}}
\put(4,0){\vector(-1,0){1}}
\put(0,0){\vector(2,3){0.5}}
\put(2,3){\vector(-2,-3){0.5}}
\put(4,0){\vector(-2,3){0.5}}
\put(4,0){\vector(-2,3){2}}
\put(2,-1){\makebox(0,0){$g(n)$}}
\put(8,0){\line(1,0){4}}
\put(8,0){\line(2,3){2}}
\put(12,0){\line(-2,3){2}}
\put(8,0){\vector(1,0){1}}
\put(8,0){\vector(1,0){4}}
\put(10,3){\vector(-2,-3){0.5}}
\put(8,0){\vector(2,3){0.5}}
\put(10,3){\vector(2,-3){2}}
\put(12,0){\vector(-2,3){2}}
\put(10,-1){\makebox(0,0){$pa(n)$}}
\put(16,0){\line(1,0){4}}
\put(16,0){\line(2,3){2}}
\put(20,0){\line(-2,3){2}}
\put(16,0){\vector(1,0){4}}
\put(16,0){\vector(1,0){1}}
\put(16,0){\vector(2,3){0.5}}
\put(16,0){\vector(2,3){2}}
\put(18,3){\vector(2,-3){0.5}}
\put(18,3){\vector(2,-3){2}}
\put(18,-1){\makebox(0,0){$pb(n)$}}
\put(24,0){\line(1,0){4}}
\put(24,0){\line(2,3){2}}
\put(28,0){\line(-2,3){2}}
\put(28,0){\vector(-1,0){1}}
\put(28,0){\vector(-1,0){4}}
\put(24,0){\vector(2,3){0.5}}
\put(26,3){\vector(-2,-3){0.5}}
\put(28,0){\vector(-2,3){2}}
\put(26,3){\vector(2,-3){2}}
\put(26,-1){\makebox(0,0){$pc(n)$}}
\put(32,0){\line(1,0){4}}
\put(32,0){\line(2,3){2}}
\put(36,0){\line(-2,3){2}}
\put(36,0){\vector(-1,0){4}}
\put(36,0){\vector(-1,0){1}}
\put(32,0){\vector(2,3){0.5}}
\put(32,0){\vector(2,3){2}}
\put(34,3){\vector(2,-3){0.5}}
\put(34,3){\vector(2,-3){2}}
\put(34,-1){\makebox(0,0){$pd(n)$}}
\put(40,0){\line(1,0){4}}
\put(40,0){\line(2,3){2}}
\put(44,0){\line(-2,3){2}}
\put(40,0){\vector(1,0){4}}
\put(44,0){\vector(-1,0){4}}
\put(42,3){\vector(-2,-3){2}}
\put(40,0){\vector(2,3){2}}
\put(42,3){\vector(2,-3){0.5}}
\put(42,3){\vector(2,-3){2}}
\put(42,-1){\makebox(0,0){$r(n)$}}
\end{picture}

\vspace*{5mm}
\caption{\footnotesize{Illustration for the configurations $g(n)$, $pa(n)$, $pb(n)$, $pc(n)$, $pd(n)$, and $r(n)$. Only the three outmost vertices and the directions of the edges connected to them are shown explicitly.}} 
\label{gprfig}
\end{figure}
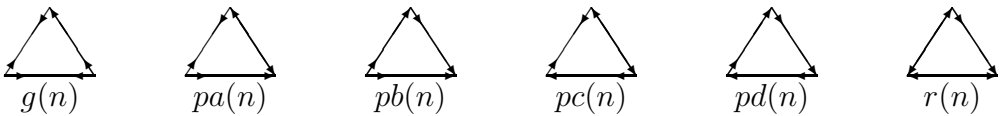

\begin{defi} \label{defisg2} Consider the generalized two-dimensional Sierpinski gasket $SG_b(n)$ at stage $n$. (i) Define $g_b(n)$ as the number of ice model configurations such that one certain edge connected to an outmost vertex, say the topmost vertex in Fig. \ref{gprfig}, is directed inward and the other five edges connected to the outmost vertices are directed outward. (ii) Define $pa_b(n)$ as the number of ice model configurations such that the two edges of a certain outmost vertex, say the left one in Fig. \ref{gprfig}, are directed outward; the two edges of another certain outmost vertex, say the right one in Fig. \ref{gprfig},  are directed inward; only one of the edges of the third outmost vertex is directed inward and it is on the same side of the Sierpinski gasket as one of the edges of the second outmost vertex. (iii) Define $pb_b(n)$ as the numbers of ice model configurations such that the two edges of a certain outmost vertex, say the left one in Fig. \ref{gprfig}, are directed outward; the two edges of another certain outmost vertex, say the right one in Fig. \ref{gprfig}, are directed inward; only one of the edges of the third outmost vertex is directed outward and it is on the same side of the Sierpinski gasket as one of the edges of the second outmost vertex. (iv) Define $pc_b(n)$ as the numbers of ice model configurations such that all three outmost vertices have one edge directed inward and one edge directed outward, while two certain directed-inward edges are on the same side of the Sierpinski gasket, say the upper-right side in Fig. \ref{gprfig}. (v) Define $pd_b(n)$ as the number of ice model configurations such that all three outmost vertices have one edge directed inward and one edge directed outward, while all three directed-inward edges are on the different sides of the Sierpinski gasket with a certain direction, say clockwise in Fig. \ref{gprfig}. (vi) Define $r_b(n)$ as the number of ice model configurations such that one certain edge connected to an outmost vertex, say the topmost vertex in Fig. \ref{gprfig}, is directed outward and the other five edges connected to the outmost vertices are directed inward.
\end{defi}

\bigskip

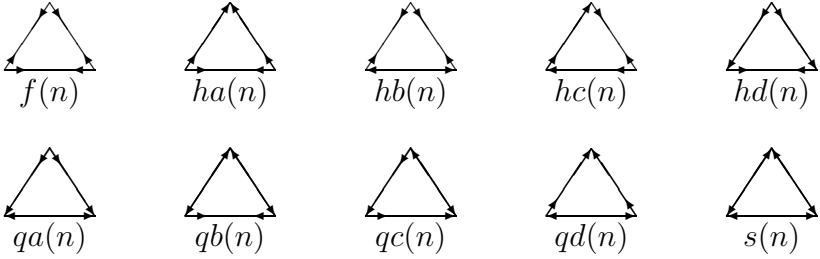
\begin{figure}[htbp]
\unitlength 3mm 
\begin{picture}(36,3)
\put(0,0){\line(1,0){4}}
\put(0,0){\line(2,3){2}}
\put(4,0){\line(-2,3){2}}
\put(0,0){\vector(1,0){1}}
\put(4,0){\vector(-1,0){1}}
\put(0,0){\vector(2,3){0.5}}
\put(2,3){\vector(-2,-3){0.5}}
\put(4,0){\vector(-2,3){0.5}}
\put(2,3){\vector(2,-3){0.5}}
\put(2,-1){\makebox(0,0){$f(n)$}}
\put(8,0){\line(1,0){4}}
\put(8,0){\line(2,3){2}}
\put(12,0){\line(-2,3){2}}
\put(8,0){\vector(1,0){1}}
\put(12,0){\vector(-1,0){1}}
\put(8,0){\vector(2,3){0.5}}
\put(8,0){\vector(2,3){2}}
\put(12,0){\vector(-2,3){0.5}}
\put(12,0){\vector(-2,3){2}}
\put(10,-1){\makebox(0,0){$ha(n)$}}
\put(16,0){\line(1,0){4}}
\put(16,0){\line(2,3){2}}
\put(20,0){\line(-2,3){2}}
\put(16,0){\vector(1,0){4}}
\put(20,0){\vector(-1,0){4}}
\put(16,0){\vector(2,3){0.5}}
\put(18,3){\vector(-2,-3){0.5}}
\put(18,3){\vector(2,-3){0.5}}
\put(20,0){\vector(-2,3){0.5}}
\put(18,-1){\makebox(0,0){$hb(n)$}}
\put(24,0){\line(1,0){4}}
\put(24,0){\line(2,3){2}}
\put(28,0){\line(-2,3){2}}
\put(28,0){\vector(-1,0){4}}
\put(28,0){\vector(-1,0){1}}
\put(24,0){\vector(2,3){0.5}}
\put(24,0){\vector(2,3){2}}
\put(28,0){\vector(-2,3){0.5}}
\put(26,3){\vector(2,-3){0.5}}
\put(26,-1){\makebox(0,0){$hc(n)$}}
\put(32,0){\line(1,0){4}}
\put(32,0){\line(2,3){2}}
\put(36,0){\line(-2,3){2}}
\put(32,0){\vector(1,0){1}}
\put(36,0){\vector(-1,0){1}}
\put(34,3){\vector(-2,-3){0.5}}
\put(34,3){\vector(-2,-3){2}}
\put(34,3){\vector(2,-3){0.5}}
\put(34,3){\vector(2,-3){2}}
\put(34,-1){\makebox(0,0){$hd(n)$}}
\end{picture}

\vspace*{10mm}
\begin{picture}(36,3)
\put(0,0){\line(1,0){4}}
\put(0,0){\line(2,3){2}}
\put(4,0){\line(-2,3){2}}
\put(0,0){\vector(1,0){4}}
\put(4,0){\vector(-1,0){4}}
\put(2,3){\vector(-2,-3){0.5}}
\put(2,3){\vector(-2,-3){2}}
\put(2,3){\vector(2,-3){0.5}}
\put(2,3){\vector(2,-3){2}}
\put(2,-1){\makebox(0,0){$qa(n)$}}
\put(8,0){\line(1,0){4}}
\put(8,0){\line(2,3){2}}
\put(12,0){\line(-2,3){2}}
\put(8,0){\vector(1,0){1}}
\put(12,0){\vector(-1,0){1}}
\put(8,0){\vector(2,3){2}}
\put(10,3){\vector(-2,-3){2}}
\put(12,0){\vector(-2,3){2}}
\put(10,3){\vector(2,-3){2}}
\put(10,-1){\makebox(0,0){$qb(n)$}}
\put(16,0){\line(1,0){4}}
\put(16,0){\line(2,3){2}}
\put(20,0){\line(-2,3){2}}
\put(16,0){\vector(1,0){1}}
\put(16,0){\vector(1,0){4}}
\put(18,3){\vector(-2,-3){0.5}}
\put(18,3){\vector(-2,-3){2}}
\put(18,3){\vector(2,-3){2}}
\put(20,0){\vector(-2,3){2}}
\put(18,-1){\makebox(0,0){$qc(n)$}}
\put(24,0){\line(1,0){4}}
\put(24,0){\line(2,3){2}}
\put(28,0){\line(-2,3){2}}
\put(24,0){\vector(1,0){4}}
\put(28,0){\vector(-1,0){4}}
\put(24,0){\vector(2,3){0.5}}
\put(24,0){\vector(2,3){2}}
\put(28,0){\vector(-2,3){0.5}}
\put(28,0){\vector(-2,3){2}}
\put(26,-1){\makebox(0,0){$qd(n)$}}
\put(32,0){\line(1,0){4}}
\put(32,0){\line(2,3){2}}
\put(36,0){\line(-2,3){2}}
\put(32,0){\vector(1,0){4}}
\put(36,0){\vector(-1,0){4}}
\put(32,0){\vector(2,3){2}}
\put(34,3){\vector(-2,-3){2}}
\put(34,3){\vector(2,-3){2}}
\put(36,0){\vector(-2,3){2}}
\put(34,-1){\makebox(0,0){$s(n)$}}
\end{picture}

\vspace*{5mm}
\caption{\footnotesize{Illustration for other possible edge directions which are not used here. Out of the six edges connected to the three outmost vertices, $f(n)$ has all of them directed outward; $ha(n)$, $hb(n)$, $hc(n)$, $hd(n)$ have four of them directed outward; $qa(n)$, $qb(n)$, $qc(n)$, $qd(n)$ have four of them directed inward; $s(n)$ has all of them directed inward.}} 
\label{fhqsfig}
\end{figure}

Since we only consider ordinary Sierpinski gasket in this and next sections, we shall use the notations $g(n)$, $pa(n)$, $pb(n)$, $pc(n)$, $pd(n)$, and $r(n)$ for simplicity. They are illustrated in Fig. \ref{gprfig}, where only the outmost vertices and the directions of the edges connected to them are shown. In principle, the edges of the three outmost vertices have other possible directions as illustrated in Fig. \ref{fhqsfig}, but they do not appear in our consideration as discuss below. Because of rotational and reflection symmetries, $g(n)$, $pa(n)$, $pb(n)$, $pc(n)$ and $r(n)$ have multiplicity six, while $pd(n)$ have multiplicity two. It is clear that the initial values at stage zero are $g(0)=pa(0)=pc(0)=r(0)=0$ and $pb(0)=pd(0)=1$. For the purpose of obtaining the asymptotic behavior of $I(n)$ in this section, $g(n)$ and $r(n)$ are not needed such that
\beq
I(n) = 6pa(n)+6pb(n)+6pc(n)+2pd(n) 
\label{isg2}
\eeq
for non-negative integer $n$. The reason of missing $g(n)$ and $r(n)$ is due to their zero value at stage zero. For example, $g(n+1)$ may contain a term like $g(n)pb^2(n)$ in its recursion relations, but all such terms are equal to zero since $g(0)=0$. However, $g(n)$ and $r(n)$ are nonzero for $n > 0$ in the next section for the eight-vertex model. Now the four quantities $pa(n)$, $pb(n)$, $pc(n)$ and $pd(n)$ satisfy recursion relations. 

\bigskip

\bigskip

\bigskip

\begin{lemma} \label{lemmasg2r} For any non-negative integer $n$,
\beq
pa(n+1) = [pa(n)+pb(n)]^2 [pa(n)+3pc(n)+pd(n)] \ , 
\label{paeq}
\eeq
\beq
pb(n+1) = [pa(n)+pb(n)]^2 [pb(n)+3pc(n)+pd(n)] \ , 
\label{pbeq}
\eeq
\beq
pc(n+1) = pa^2(n)pb(n) + pb^2(n)pa(n) + [3pc(n)+pd(n)]^3 \ , 
\label{pceq}
\eeq
\beq
pd(n+1) = pa^3(n) + pb^3(n) + [3pc(n)+pd(n)]^3 \ .
\label{pdeq}
\eeq
\end{lemma}

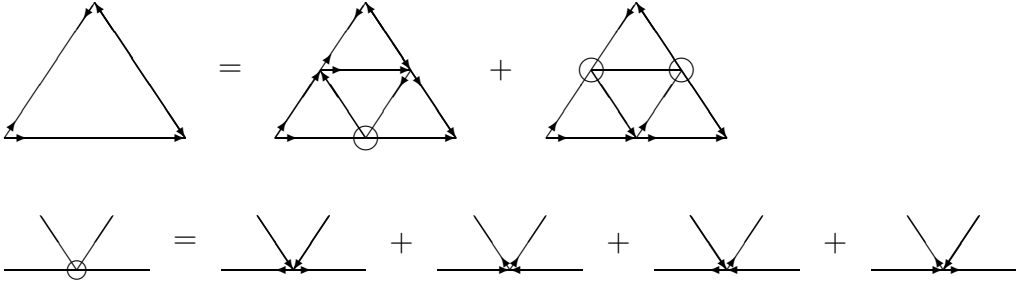
\begin{figure}[htbp]
\unitlength 3mm 
\begin{picture}(32,6)
\put(0,0){\line(1,0){8}}
\put(0,0){\line(2,3){4}}
\put(8,0){\line(-2,3){4}}
\put(0,0){\vector(1,0){1}}
\put(0,0){\vector(1,0){8}}
\put(0,0){\vector(2,3){0.5}}
\put(4,6){\vector(-2,-3){0.5}}
\put(4,6){\vector(2,-3){4}}
\put(8,0){\vector(-2,3){4}}
\put(10,3){\makebox(0,0){$=$}}
\put(12,0){\line(1,0){8}}
\put(12,0){\line(2,3){4}}
\put(20,0){\line(-2,3){4}}
\put(12,0){\vector(1,0){1}}
\put(12,0){\vector(1,0){8}}
\put(12,0){\vector(2,3){0.5}}
\put(16,6){\vector(-2,-3){0.5}}
\put(16,6){\vector(2,-3){4}}
\put(20,0){\vector(-2,3){4}}
\put(14,3){\line(1,0){4}}
\put(16,0){\line(2,3){2}}
\put(16,0){\line(-2,3){2}}
\put(14,3){\vector(1,0){1}}
\put(14,3){\vector(1,0){4}}
\put(12,0){\vector(2,3){2}}
\put(14,3){\vector(2,3){0.5}}
\put(16,0){\vector(-2,3){2}}
\put(16,6){\vector(2,-3){2}}
\put(18,3){\vector(2,-3){0.5}}
\put(18,3){\vector(-2,-3){0.5}}
\put(16,0){\circle{1}}
\put(22,3){\makebox(0,0){$+$}}
\put(24,0){\line(1,0){8}}
\put(24,0){\line(2,3){4}}
\put(32,0){\line(-2,3){4}}
\put(24,0){\vector(1,0){1}}
\put(24,0){\vector(1,0){8}}
\put(24,0){\vector(2,3){0.5}}
\put(28,6){\vector(-2,-3){0.5}}
\put(28,6){\vector(2,-3){4}}
\put(32,0){\vector(-2,3){4}}
\put(26,3){\line(1,0){4}}
\put(28,0){\line(2,3){2}}
\put(28,0){\line(-2,3){2}}
\put(24,0){\vector(1,0){4}}
\put(28,0){\vector(1,0){1}}
\put(28,0){\vector(2,3){0.5}}
\put(26,3){\vector(2,-3){2}}
\multiput(26,3)(4,0){2}{\circle{1}}
\end{picture}

\unitlength 2.4mm
\vspace*{10mm}
\begin{picture}(56,3)
\put(0,0){\line(1,0){8}}
\put(4,0){\line(2,3){2}}
\put(4,0){\line(-2,3){2}}
\put(4,0){\circle{1}}
\put(10,1.5){\makebox(0,0){$=$}}
\put(12,0){\line(1,0){8}}
\put(16,0){\line(2,3){2}}
\put(16,0){\line(-2,3){2}}
\put(16,0){\vector(1,0){1}}
\put(16,0){\vector(-1,0){1}}
\put(14,3){\vector(2,-3){2}}
\put(18,3){\vector(-2,-3){2}}
\put(22,1.5){\makebox(0,0){$+$}}
\put(24,0){\line(1,0){8}}
\put(28,0){\line(2,3){2}}
\put(28,0){\line(-2,3){2}}
\put(24,0){\vector(1,0){4}}
\put(32,0){\vector(-1,0){4}}
\put(28,0){\vector(-2,3){0.5}}
\put(28,0){\vector(2,3){0.5}}
\put(34,1.5){\makebox(0,0){$+$}}
\put(36,0){\line(1,0){8}}
\put(40,0){\line(2,3){2}}
\put(40,0){\line(-2,3){2}}
\put(40,0){\vector(-1,0){1}}
\put(44,0){\vector(-1,0){4}}
\put(38,3){\vector(2,-3){2}}
\put(40,0){\vector(2,3){0.5}}
\put(46,1.5){\makebox(0,0){$+$}}
\put(48,0){\line(1,0){8}}
\put(52,0){\line(2,3){2}}
\put(52,0){\line(-2,3){2}}
\put(48,0){\vector(1,0){4}}
\put(52,0){\vector(1,0){1}}
\put(52,0){\vector(-2,3){0.5}}
\put(54,3){\vector(-2,-3){2}}
\end{picture}

\caption{\footnotesize{Illustration for the expression of $pa(n+1)$. The representation of a big circle at an identified vertex corresponds to four possible configurations such that for each $SG(n)$ one edge is directed inward and the other edge is outward.}} 
\label{pafig}
\end{figure}

{\sl Proof} \quad 
The Sierpinski gaskets $SG(n+1)$ is composed of three $SG(n)$ with three pairs of vertices identified. At these identified vertices, the ice rule should be satisfied.
As illustrated in Fig. \ref{pafig}, the number $pa(n+1)$ consists of two cases. For the first case, the top $SG(n)$ is always counted by $pa(n)$. The identified vertex of the lower two $SG(n)$'s has four possible configurations such that for each of these $SG(n)$'s one edge is directed inward and the other edge is outward as represented by a big circle in Fig. \ref{pafig}. They are counted by $pa(n) [pa(n)+pb(n)]^2$. One may wonder why the other two possible configurations satisfying the ice rule are missing. If the two edges of the left $SG(n)$ are directed outward and the two edges of the right $SG(n)$ are directed inward, then the left $SG(n)$ is counted by $ha(n)$ and the right $SG(n)$ is counted by $qa(n)$. However, such terms have no contributions because $ha(n)$, $qa(n)$ are always zero for any $n$ by the same reason discussed earlier. For the second case, the directions of the edges of the bottom identified vertex is fixed as shown in Fig. \ref{pafig}. There are four possible configurations for the upper-left identified vertex multiplying four possible configurations for the upper-right identified vertex, and they are counted by $[pa(n)+pb(n)]^2 [3pc(n)+pd(n)]$. Eq. (\ref{paeq}) is verified by combining the results of these two cases.

\bigskip

\bigskip

The number $pb(n+1)$ is almost the same as $pa(n+1)$ except that the two edges of the topmost vertex change directions. It follows that the top $SG(n)$ is always counted by $pb(n)$ for the first case, while the results for the second case remain the same, so that Eq. (\ref{pbeq}) is verified.

\begin{figure}[htbp]
\unitlength 3mm 
\begin{picture}(44,6)
\put(0,0){\line(1,0){8}}
\put(0,0){\line(2,3){4}}
\put(8,0){\line(-2,3){4}}
\put(8,0){\vector(-1,0){1}}
\put(8,0){\vector(-1,0){8}}
\put(0,0){\vector(2,3){0.5}}
\put(4,6){\vector(-2,-3){0.5}}
\put(4,6){\vector(2,-3){4}}
\put(8,0){\vector(-2,3){4}}
\put(10,3){\makebox(0,0){$=$}}
\put(12,0){\line(1,0){8}}
\put(12,0){\line(2,3){4}}
\put(20,0){\line(-2,3){4}}
\put(20,0){\vector(-1,0){1}}
\put(20,0){\vector(-1,0){8}}
\put(12,0){\vector(2,3){0.5}}
\put(16,6){\vector(-2,-3){0.5}}
\put(16,6){\vector(2,-3){4}}
\put(20,0){\vector(-2,3){4}}
\put(14,3){\line(1,0){4}}
\put(16,0){\line(2,3){2}}
\put(16,0){\line(-2,3){2}}
\put(18,3){\vector(-1,0){1}}
\put(18,3){\vector(-1,0){4}}
\put(14,3){\vector(2,-3){0.5}}
\put(14,3){\vector(-2,-3){0.5}}
\put(16,6){\vector(-2,-3){2}}
\put(16,0){\vector(2,3){2}}
\put(20,0){\vector(-2,3){2}}
\put(18,3){\vector(-2,3){0.5}}
\put(12,0){\vector(1,0){4}}
\put(16,0){\vector(1,0){1}}
\put(16,0){\vector(2,3){0.5}}
\put(14,3){\vector(2,-3){2}}
\put(22,3){\makebox(0,0){$+$}}
\put(24,0){\line(1,0){8}}
\put(24,0){\line(2,3){4}}
\put(32,0){\line(-2,3){4}}
\put(32,0){\vector(-1,0){1}}
\put(32,0){\vector(-1,0){8}}
\put(24,0){\vector(2,3){0.5}}
\put(28,6){\vector(-2,-3){0.5}}
\put(28,6){\vector(2,-3){4}}
\put(32,0){\vector(-2,3){4}}
\put(26,3){\line(1,0){4}}
\put(28,0){\line(2,3){2}}
\put(28,0){\line(-2,3){2}}
\put(26,3){\vector(1,0){1}}
\put(26,3){\vector(1,0){4}}
\put(24,0){\vector(2,3){2}}
\put(26,3){\vector(2,3){0.5}}
\put(28,0){\vector(-2,3){2}}
\put(28,6){\vector(2,-3){2}}
\put(30,3){\vector(2,-3){0.5}}
\put(30,3){\vector(-2,-3){0.5}}
\put(32,0){\vector(-1,0){4}}
\put(28,0){\vector(-1,0){1}}
\put(28,0){\vector(-2,3){0.5}}
\put(30,3){\vector(-2,-3){2}}
\put(34,3){\makebox(0,0){$+$}}
\put(36,0){\line(1,0){8}}
\put(36,0){\line(2,3){4}}
\put(44,0){\line(-2,3){4}}
\put(44,0){\vector(-1,0){1}}
\put(44,0){\vector(-1,0){8}}
\put(36,0){\vector(2,3){0.5}}
\put(40,6){\vector(-2,-3){0.5}}
\put(40,6){\vector(2,-3){4}}
\put(44,0){\vector(-2,3){4}}
\put(38,3){\line(1,0){4}}
\put(40,0){\line(2,3){2}}
\put(40,0){\line(-2,3){2}}
\put(40,0){\circle{1}}
\multiput(38,3)(4,0){2}{\circle{1}}
\end{picture}

\caption{\footnotesize{Illustration for the expression of $pc(n+1)$.}} 
\label{pcfig}
\end{figure}
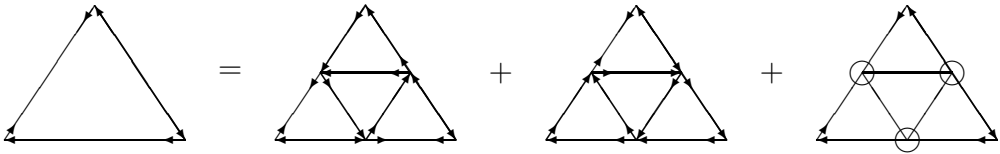

As illustrated in Fig. \ref{pcfig}, the number $pc(n+1)$ consists of three cases. For the first case, the directions of the edges of the three identified vertices are fixed and the number is counted by $pa^2(n)pb(n)$. Reversing the directions of the edges of the three identified vertices in the first case gives the second case, which is counted by $pb^2(n)pa(n)$. Finally for the third case, each of the three identified vertex has four possible configurations such that for each related $SG(n)$ one edge is directed inward and the other is outward. This number if counted by $[3pc(n)+pd(n)]^3$, and Eq. (\ref{pceq}) is verified.

\bigskip

The number $pd(n+1)$ is almost the same as $pc(n+1)$ except that the two edges of the topmost vertex change directions. It follows that the first case is counted by $pa^3(n)$, the second case is counted by $pb^3(n)$, while the results for the third case remain the same, so that Eq. (\ref{pdeq}) is verified.
\ $\Box$

\bigskip

The values of $pa(n)$, $pb(n)$, $pc(n)$, $pd(n)$, $I(n)$ for small $n$ can be evaluated recursively by Eqs. (\ref{paeq})-(\ref{pdeq}) as listed in Table \ref{tablesg2}. These numbers grow exponentially, and do not have simple integer factorizations. To estimate the value of the entropy defined in Eq. (\ref{zdef}), we need the following lemmas. 
\bigskip

\begin{table}[htbp]
\caption{\label{tablesg2} The first few values of $pa(n)$, $pb(n)$, $pc(n)$, $pd(n)$, $I(n)$.}
\begin{center}
\begin{tabular}{|c||r|r|r|r|r|}
\hline\hline 
$n$     & 0 &  1 &     2 &             3 & 4 \\ \hline\hline 
$pa(n)$ & 0 &  1 &    54 &     7,953,309 & 152,890,249,552,106,555,312,694 \\ \hline 
$pb(n)$ & 1 &  2 &    63 &     8,076,510 & 152,921,906,677,033,336,640,655 \\ \hline 
$pc(n)$ & 0 &  1 &   131 &   146,761,217 & 202,319,214,683,073,568,675,255,835 \\ \hline 
$pd(n)$ & 1 &  2 &   134 &   146,770,694 & 202,319,214,926,381,958,377,247,254 \\ \hline 
$I(n)$  & 8 & 28 & 1,756 & 1,270,287,604 & 1,620,388,590,888,580,168,157,749,612 \\  \hline\hline 
\end{tabular}
\end{center}
\end{table}

\bigskip

Let us use the notations $\alpha(n) = pa(n)/pb(n)$,  $\beta(n) = pd(n)/pc(n)$ and $\gamma(n) = pb(n)/pc(n)$ for $n \geq 1$.
 
\begin{lemma} \label{lemmasg2abc} $\alpha(n) \in (0,1)$ and $\beta(n) > 1$ for any positive integer $n$. Furthermore,
\beq
0 < 1 - \alpha(n+1) < \frac{\gamma(n)}{4} \bigl[ 1-\alpha(n) \bigr] \ , 
\eeq
\beq
\frac{\alpha(n) \gamma(n)^3 [1-\alpha(n)]^2 } {32\beta(n)^3 + \gamma(n)^3} < \beta(n+1)-1 < \frac{\gamma(n)^3 [1-\alpha(n)]^2 } {32} \ ,
\eeq
\beq
\frac{4\gamma(n)^2 \bigl[ 4+\gamma(n) \bigr]} {\gamma(n)^3 [1+\alpha(n)^{-1}] + 64\alpha(n)^{-2}\beta(n)} < \gamma(n+1) < \frac{\gamma(n)^2 \bigl[ 3+\gamma(n)+\beta(n) \bigr]} {16} 
\label{ratiosg2}
\eeq
for $n \geq 1$, such that the sequence $\alpha(n)$ increases to one, $\beta(n)$ decreases to one, and $\gamma(n)$ decreases to zero as $n$ increases.

\end{lemma}

{\sl Proof} \quad 
By Eqs. (\ref{paeq}) and (\ref{pbeq}), we have
\beq
pb(n+1) - pa(n+1) = [pb(n)-pa(n)] [pa(n)+pb(n)]^2 \ ,
\eeq
and $pa(n) < pb(n)$ is established by mathematical induction hypothesis $\alpha(n) \in (0,1)$ and the initial value $\alpha(1) = 1/2$. By Eqs. (\ref{pceq}) and (\ref{pdeq}), we have
\beq
pd(n+1) - pc(n+1) = [pa(n)+pb(n)] [pb(n)-pa(n)]^2 > 0 \ ,
\eeq
so that $\beta(n) > 1$ for all $n \geq 1$.

Next using the fact that $pa(n) < pb(n)$ and $pc(n) < pd(n)$, we have
\beq
0 < 1-\alpha(n+1) = 1 - \frac{pa(n)+3pc(n)+pd(n)} {pb(n)+3pc(n)+pd(n)} < \frac{pb(n)-pa(n)}{4pc(n)} = \frac{\gamma(n)}{4} \bigl[1-\alpha(n) \bigr] \ ,
\label{alpha}
\eeq 
and
\beq
\beta(n+1)-1 = \frac{[pa(n)+pb(n)][pb(n)-pa(n)]^2} {pa(n)pb(n)[pa(n)+pb(n)]+[3pc(n)+pd(n)]^3} \ ,
\eeq
such that
\beqs
& & \frac{\alpha(n) \gamma(n)^3 [1-\alpha(n)]^2 } {32\beta(n)^3 + \gamma(n)^3} = \frac{2pa(n)[pb(n)-pa(n)]^2}{2pb(n)^3+[4pd(n)]^3}  < \beta(n+1)-1 \cr\cr
& & < \frac{2pb(n)[pb(n)-pa(n)]^2}{[4pc(n)]^3} = \frac{\gamma(n)^3[1-\alpha(n)]^2}{32} \ .
\label{beta}
\eeqs
Finial, by Eqs. (\ref{pbeq}) and (\ref{pceq}), we have
\beq
\gamma(n+1) =  \gamma(n)^2\times\frac { [\alpha(n)+1]^2[\gamma(n)+3+\beta(n)]}{\gamma(n)^3\alpha(n)^2[1+\alpha(n)^{-1}]+[3+\beta(n)]^3} \ ,
\eeq
so that
\beq
\frac{4\alpha(n)^2 \gamma(n)^2 \bigl[ 4+\gamma(n) \bigr]} {\gamma(n)^3\alpha(n)^2 [1+\alpha(n)^{-1}] + 64\beta(n)^{3}} < \gamma(n+1) < \frac{ \gamma(n)^2 \bigl[ 3+\gamma(n)+\beta(n) \bigr]} {16} \ .
\label{gamma}
\eeq
By Eqs. (\ref{alpha}), (\ref{beta}) and (\ref{gamma}), we have
\beqs
& & \frac{1-\alpha(n+1)} {1-\alpha(n)} < \frac{\gamma(n)}{4} \ , \qquad
\frac{\gamma(n+1)} {\gamma(n)} < \frac{ \gamma(n) \bigl[ 3+\gamma(n)+\beta(n) \bigr]} {16} \ , \cr\cr
& & \frac{\beta(n+2)-1}{\beta(n+1)-1} < \frac{\gamma(n+1)^3[1-\alpha(n+1)]^2}{32[\beta(n+1)-1]} < \frac{\gamma(n)^5 [3+\gamma(n)+\beta(n)]^3 [32\beta(n)^3+\gamma(n)^3]}{2^{21}\alpha(n)} \ . \cr
& &
\eeqs
With $\alpha(1)=1/2$, $\beta(1)=2$ and $\gamma(1)=2$, it is easy to see that $\alpha(n)$ increases to one, $\beta(n)$ decreases to one and $\gamma(n)$ decreases to zero as $n$ increases. 
\ $\Box$

\bigskip

In passing, we notice that $\gamma(n) / [1-\alpha(n)] \rightarrow (0,\infty)$ and $[\beta(n)-1]/\gamma(n)^{5/2} \rightarrow (0,\infty)$ in the infinite $n$ limit.

\bigskip

\begin{lemma} \label{lemmasg2b} The entropy for the number of ice model configurations on $SG(n)$ is bounded:
\beq
\frac{2}{3^{m+1}} \ln pc(m) + \frac{2}{3^m} \ln 2 \le S_{I,SG} \le \frac{2}{3^{m+1}} \ln pd(m) + \frac{2}{3^m} \ln 2 \ ,
\label{zsg2}
\eeq
where $m$ is a positive integer.
\end{lemma}

{\sl Proof} \quad 
By Eq. (\ref{pceq}) and Lemma \ref{lemmasg2abc}, we have
\beq
pc(n) > 64 pc^3(n-1) > 64 [64pc^3(n-2)]^3 > ... > pc(m)^{3^{n-m}} \times 64^{(3^{n-m}-1)/2}
\eeq
for any $m < n$. Similarly by Eq. (\ref{pdeq}) and Lemma \ref{lemmasg2abc}, we obtain
\beq
pd(n) < 64 pd^3(n-1) < 64 [64pd^3(n-2)]^3 < ... < pd(m)^{3^{n-m}} \times 64^{(3^{n-m}-1)/2} \ ,
\eeq
so that
\beq
3^{n-m} \ln pc(m) + 3(3^{n-m}-1) \ln 2 < \ln pc(n) < \ln pd(n) < 3^{n-m} \ln pd(m) + 3(3^{n-m}-1) \ln 2 \ .
\label{pdbound}
\eeq

By Eqs. (\ref{v}) and (\ref{isg2}), we have
\beq
\frac {\ln I(n)}{v(SG(n))} = \frac {2}{3(3^n+1)} \ln \bigl [ 2 + 6\frac{pc(n)}{pd(n)} + 6\frac{pb(n)}{pd(n)} + 6\frac{pa(n)}{pd(n)} \bigr ] + \frac {2\ln pd(n)}{3(3^n+1)} \ .
\eeq
By the definition of the entropy in Eq. (\ref{zdef}) and Lemma \ref{lemmasg2abc},
\beqs
S_{I,SG} & = & \lim_{n \to \infty} \frac{\ln I(n)}{v(SG(n))} \cr\cr
& = & \lim_{n\to\infty}
\frac {2}{3(3^n+1)} \ln \bigl [ 2 + 6\frac{pc(n)}{pd(n)} + 6\frac{pb(n)}{pd(n)} + 6\frac{pa(n)}{pd(n)} \bigr ] + \lim_{n\to\infty} \frac{ 2 \ln pd(n)}{3(3^n+1)} \cr\cr
& = & \lim_{n\to\infty} \frac{ 2 \ln pd(n)}{3(3^n+1)} \ .
\eeqs
The proof is completed using the inequality (\ref{pdbound}).
\ $\Box$

\bigskip

The difference between the upper and lower bounds for $S_{I,SG}$ quickly converges to zero as $m$ increases, and we have the following proposition.

\bigskip

\begin{propo} \label{proposg2} The entropy per site for the number of ice model configurations on the two-dimensional Sierpinski gasket $SG(n)$ in the large $n$ limit is $S_{I,SG}=0.515648810655...$. 

\end{propo}

\bigskip

The numerical value of $S_{I,SG}$ can be calculated with more than a hundred significant figures accurate when $m$ in Eq. (\ref{zsg2}) is equal to eight. It is too lengthy to be included here and is available from the authors on request. As the square lattice (sq) also has degree 4, it is interesting to compare our result with Lieb's on the square lattice \cite{lieb67,lieb67n}, $S_{I,sq} = \frac32 \ln \frac43 = 0.431523108677...$. We see that the entropy per site on the two-dimensional Sierpinski gasket is significantly larger.

The upper bound given in Lemma \ref{lemmasg2b} can be improved further as follows.

\bigskip

\begin{lemma} \label{lemmasg2bn} For any integers $n > m \ge 1$, 
\beqs
& & 64^{(3^{n-m}-1)/2} pc(m)^{3^{n-m}} < pc(n) \cr\cr
& < & 64^{(3^{n-m}-1)/2} pc(m)^{3^{n-m}} \bigl[ \frac{\gamma(m)^3}{32} + \bigl( 1+\frac{\beta(m)-1}{4} \bigr)^3 \bigr]^{(3^{n-m}-1)/2}  \ .
\label{pc}
\eeqs
\end{lemma}

{\sl Proof} \quad 
By Eq. (\ref{pceq}), we know
\beq
pc(n+1) = pc(n)^3 \Bigl[ \alpha(n)^2 \gamma(n)^3 + \alpha(n)\gamma(n)^3 + \bigl(3+\beta(n) \bigr)^3 \Bigr] 
\eeq
for any $n \geq 1$. By Lemma~\ref{lemmasg2abc}, we have for any positive integers $n > m \ge 1$
\beq
64pc(n-1)^3 < pc(n) < 64pc(n-1)^3 \bigl[ \frac{\gamma(m)^3}{32} + \bigl( 1+\frac{\beta(m)-1}{4} \bigr)^3 \bigr] \ .
\eeq
Using the formula $a_n = c^{(3^{n-m}-1)/2} a_m^{3^{n-m}}$ if $a_{j+1} = c a_j^3$ for $j=m$, $m+1$, ..., $n-1$, where $c$ is a constant, then Eq. (\ref{pc}) is established.
\ $\Box$

\bigskip

\begin{lemma} \label{lemmasg2d} The logarithm of the number of ice model configurations divided by the number of vertices on $SG(n)$ is bounded:
\beqs
& & \frac{2\ln 2}{3^n+1} + \frac{\bigl( 3^{-m}-3^{-n} \bigr) \ln 4}{1+3^{-n}} + \frac23 \times \frac{3^{-m}}{1+3^{-n}} \ln pc(m) \cr\cr
& < & \frac{\ln I(n)}{v(SG(n))} < \frac{2}{3^n+1} \Bigl\{ \ln 2 + \frac13 \ln \Bigl[ \beta(m) + \frac32\gamma(m) \Bigr] \Bigr\} \cr\cr
& & + \frac{3^{-m}-3^{-n}}{1+3^{-n}} \Bigl\{ \ln 4 + \ln \Bigl[ 1 + \frac{\beta(m)-1}{4} \Bigr] + \frac13 \ln \Bigl[ 1+\frac{\gamma(m)^3}{32} \Bigr] \Bigr\} + \frac23 \times \frac{3^{-m}}{1+3^{-n}} \ln pc(m) \cr
& & 
\label{finiten}
\eeqs
for positive integers $n > m \ge 1$.

\end{lemma}

{\sl Proof} \quad 
The number of ice model configurations can be rewritten as
\beqs
I(n) & = & 6 \bigl[ pa(n)+pb(n)+pc(n) \bigr] + 2pd(n) \cr\cr
& = & 2pc(n) \Bigl\{ 3 \bigl[ \alpha(n)+1 \bigr] \gamma(n) + 3 + \beta(n) \Bigr\} \ ,
\eeqs
then by Lemma~\ref{lemmasg2abc}, 
\beq\label{In}
8pc(n) \Bigl[ 1+\frac32\alpha(n)\gamma(n) \Bigr] < I(n) < 8pc(n) \Bigl[ \beta(n) + \frac32\gamma(n) \Bigr] \ .
\eeq
It follows that
\beqs
& & \frac{2}{3(3^n+1)} \Bigl[ 3\ln 2 + \ln pc(n) \Bigr] < \frac{\ln I(n)}{v(SG(n))} \cr\cr
& < & \frac{2}{3(3^n+1)} \Bigl\{ 3\ln 2 + \ln pc(n) + \ln \bigl[ \beta(m) + \frac32\gamma(m) \bigr] \Bigr\} \ .
\label{ivb}
\eeqs
By Lemma~\ref{lemmasg2bn}, the upper and lower bounds of $\ln pc(n)$ are given by
\beqs
& & \frac{3^{n-m}-1}{2} \bigl( 3\ln 4 \bigr) + 3^{n-m} \ln pc(m) < \ln pc(n) \cr\cr
& < & \frac{3^{n-m}-1}{2} \Bigl\{ 3\ln 4 + \ln \Bigl[ \frac{\gamma(m)^3}{32} + \bigl( 1+\frac{\beta(m)-1}{4} \bigr)^3 \Bigr] \Bigr\} + 3^{n-m} \ln pc(m) \cr\cr
& < & \frac{3^{n-m}-1}{2} \Bigl\{ 3\ln 4 + 3 \ln \Bigl[ 1+\frac{\beta(m)-1}{4} \Bigr] + \ln \Bigl[ 1+\frac{\gamma(m)^3}{32} \Bigr] \Bigr\} + 3^{n-m} \ln pc(m) \ . \cr & &
\label{pcb}
\eeqs
Combining Eqs. (\ref{ivb}) and (\ref{pcb}), the proof is completed.
\ $\Box$

Taking the limit $n \to \infty$ in Eq. (\ref{finiten}), the entropy is bounded as 
\beqs 
& & 0 < S_{I,SG} - 3^{-m} \Bigl[ \ln 4 + \frac23 \ln pc(m) \Bigr] \cr\cr
& & < 3^{-m} \Bigl\{ \ln \Bigl[ 1+\frac{\beta(m)-1}{4} \Bigr] + \frac13 \ln \Bigl[ 1+\frac{\gamma(m)^3}{32} \Bigr] \Bigr\} \ .
\eeqs
Using the inequality $\ln (1+\epsilon) < \epsilon$ for any $\epsilon \in (0,1)$, we obtain the improved upper bound
\beq 
0< S_{I,SG} - 3^{-m} \Bigl[ \ln 4 + \frac23 \ln pc(m) \Bigr] < 3^{-m} \Bigl[ \frac{\beta(m)-1}{4} + \frac{\gamma(m)^3}{96} \Bigr] \ .
\eeq

According to Lemma \ref{lemmasg2abc}, it is clear that the upper bound converges to zero quickly as $m$ increases.

\section{The number of eight-vertex model configurations on $SG(n)$} 
\label{sectionIV}

In this section, we derive the number of eight-vertex model configurations, denoted as $E(n)$, on $SG(n)$ and the entropy per site. Now directions of the four edges of a vertex (except the three outmost vertices) are allowed to be all inward or all outward. The configurations $g(n)$ and $r(n)$ shown in Fig. \ref{gprfig} are involved as their recursion relations contain nonzero terms such as $pb^3(n-1)$. We have
\beq
E(n) = 6g(n)+6pa(n)+6pb(n)+6pc(n)+2pd(n)+6r(n)
\label{esg2}
\eeq
for non-negative integer $n$. The recursion relations are lengthy and given in the appendix. As the initial values at stage zero are $g(0)=pa(0)=pc(0)=r(0)=0$ and $pb(0)=pd(0)=1$, it turns out that their values are the same for any $n > 0$:
\beq
g(n)=pa(n)=pb(n)=pc(n)=pd(n)=r(n)=2^{(3^{n+1}-7)/2} \ .
\eeq

We have the following proposition.

\bigskip

\begin{propo} \label{proposg2ev} The number of eight-vertex model configurations on the two-dimensional Sierpinski gasket $SG(n)$ is $E(n)=2^{3(3^n+1)/2}$ and the entropy per site in the large $n$ limit is $S_{E,SG} = \ln 2$.

\end{propo}

\bigskip

Compare again with the square lattice (sq). The special case of Baxter's result \cite{baxterbook} with Boltzmann weights equal to one also gives $S_{E,sq} = \ln 2$. Namely, the entropies per site for the two-dimensional Sierpinski gasket and the square lattice are the same.

\section{The number of ice model configurations on $SG_b(n)$ with $b=3$} 
\label{sectionV}

In this section, we consider the generalized two-dimensional Sierpinski gasket $SG_b(n)$ with the number of layers $b$ equal to three. 
For $SG_3(n)$, the numbers of edges and vertices are given by 
\beq
e(SG_3(n)) = 3 \times 6^n \ ,
\label{esg23}
\eeq
\beq
v(SG_3(n)) = \frac{7 \times 6^n + 8}{5} \ ,
\label{vsg23}
\eeq
where the three outmost vertices have degree two. There are $(6^n-1)/5$ vertices of $SG_3(n)$ with degree six and $6(6^n-1)/5$ vertices with degree four. For each of the vertices with degree six, we allows three edges directed inward and the other three directed outward, just like the consideration in 20-vertex triangular ice-rule problem \cite{baxter69}.

By Definition \ref{defisg2}, the number of ice model configurations is $I_3(n) = 6pa_3(n)+6pb_3(n)+6pc_3(n)+2pd_3(n)$. The initial values are the same as for $SG$: $pa_3(0)=pc_3(0)=0$ and $pb_3(0)=pd_3(0)=1$. Again, $g_3(n)$ and $r_3(n)$ are zero for any nonnegative $n$. The recursion relations are lengthy and given in the appendix.
Some values of $pa_3(n)$, $pb_3(n)$, $pc_3(n)$, $pd_3(n)$, $I_3(n)$ are listed in Table \ref{tablesg23}. These numbers grow exponentially, and do not have simple integer factorizations.

\bigskip

\begin{table}[htbp]
\caption{\label{tablesg23} The first few values of $pa_3(n)$, $pb_3(n)$, $pc_3(n)$, $pd_3(n)$, $I_3(n)$.}
\begin{center}
\begin{tabular}{|c||r|r|r|r|}
\hline\hline 
$n$       & 0 &   1 &                 2 & 3 \\ \hline\hline 
$pa_3(n)$ & 0 &  11 &    29,665,405,536 & {\tiny 8,329,624,787,357,979,293,987,412,541,852,867,738,187,867,580,946,289,512,122,074,041,155,584} \\  \hline 
$pb_3(n)$ & 1 &  15 &    29,990,772,448 & {\tiny 8,329,642,677,826,723,066,417,765,699,958,803,959,673,796,982,684,471,740,765,855,746,621,440} \\  \hline 
$pc_3(n)$ & 0 &  15 &   527,746,306,872 & {\tiny 708,045,663,245,136,812,838,888,349,048,042,396,698,172,710,195,432,765,544,715,130,291,741,523,968} \\ \hline 
$pd_3(n)$ & 1 &  18 &   527,789,051,704 & {\tiny 708,045,663,245,233,047,276,563,406,582,556,247,659,516,105,436,007,524,358,474,481,355,710,267,392} \\ \hline 
$I_3(n) $ & 8 & 282 & 4,579,993,012,544 & {\tiny 5,664,465,261,566,078,079,800,619,338,522,817,745,538,255,642,031,993,426,552,757,072,040,596,340,736} \\
\hline\hline 
\end{tabular}
\end{center}
\end{table}

\bigskip

By a similar argument as Lemma \ref{lemmasg2b}, the entropy for the number of ice model configurations on $SG_3(n)$ is bounded:
\beq
\frac{5}{7 \times 6^m} \ln pc_3(m) + \frac{15}{7 \times 6^m} \ln 2 \le S_{I,SG_3} \le \frac{5}{7 \times 6^m} \ln pd_3(m) + \frac{15}{7 \times 6^m} \ln 2 \ ,
\label{zsg23}
\eeq
where $m$ is a positive integer. We have the following proposition.

\bigskip

\begin{propo} \label{proposg23} The entropy per site for the number of ice model configurations on the generalized two-dimensional Sierpinski gasket $SG_3(n)$ in the large $n$ limit is $S_{I,SG_3}=0.576812423363...$.

\end{propo}

\bigskip

The convergence of the upper and lower bounds remains quick. More than a hundred significant figures for $S_{I,SG_3}$ can be obtained when $m$ in Eq. (\ref{zsg23}) is equal to five.

\section{The number of generalized vertex model configurations on $SG_b(n)$ with $b=3$} 
\label{sectionVI}

In this section, we derive the number of generalized vertex model configurations, denoted as $E_3(n)$, on $SG_3(n)$ and the entropy per site. For the vertices with degree four, the number of arrows pointing inward is even just like the eight-vertex model on the square lattice; while for the vertices with degree six, the number of arrows pointing inward is odd just like the 32-vertex model on the triangular lattice. The configurations $g_3(n)$ and $r_3(n)$ shown in Fig. \ref{gprfig} are again involved. We have
\beq
E_3(n) = 6g_3(n)+6pa_3(n)+6pb_3(n)+6pc_3(n)+2pd_3(n)+6r_3(n)
\label{esg3}
\eeq
for non-negative integer $n$. The recursion relations are too lengthy to be included here. They are available from the authors on request. The initial values are the same as for SG: $g_3(0)=pa_3(0)=pc_3(0)=r_3(0)=0$ and $pb_3(0)=pd_3(0)=1$. It turns out that their values are again the same for any $n > 0$:
\beq
g_3(n)=pa_3(n)=pb_3(n)=pc_3(n)=pd_3(n)=r_3(n)=2^{(8 \times 6^n-18)/5} \ .
\eeq

We have the following proposition.

\bigskip

\begin{propo} \label{proposg3ev} The number of generalized vertex model configurations on the generalized two-dimensional Sierpinski gasket $SG_3(n)$ is $E_3(n)=2^{(8 \times 6^n+7)/5}$ and the entropy per site in the large $n$ limit is $S_{E,SG_3} = \frac87 \ln 2$.

\end{propo}

\bigskip

From Propositions \ref{proposg2ev} and \ref{proposg3ev}, we conjecture that the entropy per site $S_{E,SG_b}$ for the generalized vertex model on the generalized two-dimensional Sierpinski gasket $SG_b(n)$ is equal to $c(b) \ln 2$, where the constant $c(b)$ depends on $b$ and is larger than one for $b>2$.

\section{Discussion of ice model configurations}
\label{sectionVII}

For the generalized two-dimensional Sierpinski gasket $SG_b(n)$, the numbers of edges and vertices are given by 
\beq
e(SG_b(n)) = 3 \Big [\frac{b(b+1)}{2} \Big ]^n \ ,
\label{be}
\eeq
\beq
v(SG_b(n)) = \frac{b+4}{b+2} \Big [\frac{b(b+1)}{2} \Big ]^n + \frac{2(b+1)}{b+2} \ .
\label{bv}
\eeq
The bounds of the entropies for the ice model on $SG(n)$ and $SG_3(n)$ given in sections \ref{sectionIII} and \ref{sectionV} lead to the following conjecture for general $SG_b(n)$.

\bigskip

\begin{conj} \label{conjsgd} The entropy per site for the number of ice model configurations on the generalized two-dimensional Sierpinski gasket $SG_b$ is bounded:
\beq
\frac{b+2}{(b+4)[\frac{b(b+1)}{2}]^m} \bigl [ \ln pc_b(m) + 3 \ln 2 \bigr ] \le S_{I,SG_b} \le \frac{b+2}{(b+4)[\frac{b(b+1)}{2}]^m} \bigl [ \ln pd_b(m) + 3 \ln 2 \bigr ]   \ ,
\label{zsgb}
\eeq
where $m$ is a positive integer.
\end{conj}

\bigskip

However, to calculate $pc_b(m)$ and $pd_b(m)$ for general $b$ may be difficult.
We notice that $S_{I,SG_3}$ is a bit larger than $S_{I,SG}$. It is expected that the value of $S_{I,SG_b}$ increases slightly as $b$ increases for the generalized two-dimensional Sierpinski gasket.

\section*{acknowledgments}
The research of S.C.C. was partially supported by the NSC grants NSC-100-2112-M-006-003-MY3 and NSC-100-2119-M-002-001. The research of L.C.C was partially supported by the NSC grant NSC-99-2115-M-030-004-MY3.

\appendix

\section{Recursion relations for the eight-vertex model on $SG(n)$}

We give the recursion relations for the eight-vertex model on the two-dimensional Sierpinski gasket $SG(n)$ here. We will use the simplified notation $g_{n+1}$ to denote $g(n+1)$ and similar notations for other quantities. For any non-negative integer $n$, we have
\beqs
g_{n+1} & = & 4g_n^3 + 8g_n^2pa_n + 8g_n^2pb_n + 12g_n^2pc_n + 4g_n^2pd_n + 5g_npa_n^2 + 5g_npb_n^2 + 10g_npa_npb_n \cr
& & + 12g_npa_npc_n + 4g_npa_npd_n + 12g_npb_npc_n + 4g_npb_npd_n + 4g_n^2r_n + pa_n^3 + pb_n^3 \cr
& & + 3pa_n^2pb_n + 3pa_n^2pc_n + pa_n^2pd_n + 3pb_n^2pa_n + 3pb_n^2pc_n + pb_n^2pd_n + 6pa_npb_npc_n \cr
& & + 2pa_npb_npd_n + 4g_npa_nr_n + 4g_npb_nr_n + pa_n^2r_n + pb_n^2r_n + 2pa_npb_nr_n
\ ,
\label{geveq}
\eeqs
\beqs
\lefteqn{pa_{n+1} = pb_{n+1}} \cr
& = & 2g_n^2pa_n + 2g_n^2pb_n + 3g_npa_n^2 + 3g_npb_n^2 + 6g_npa_npb_n + 6g_npa_npc_n + 2g_npa_npd_n + 6g_npb_npc_n \cr
& & + 2g_npb_npd_n + 4g_n^2r_n + pa_n^3 + pb_n^3 + 3pa_n^2pb_n + 3pa_n^2pc_n + pa_n^2pd_n + 3pb_n^2pa_n + 3pb_n^2pc_n \cr
& & + pb_n^2pd_n + 6pa_npb_npc_n + 2pa_npb_npd_n + 8g_npa_nr_n + 8g_npb_nr_n + 12g_npc_nr_n + 4g_npd_nr_n \cr
& & + 3pa_n^2r_n + 3pb_n^2r_n + 6pa_npb_nr_n + 6pa_npc_nr_n + 2pa_npd_nr_n + 6pb_npc_nr_n + 2pb_npd_nr_n \cr
& & + 4g_nr_n^2 + 2pa_nr_n^2 + 2pb_nr_n^2
\ ,
\label{papbeveq}
\eeqs
\beqs
\lefteqn{pc_{n+1} = pd_{n+1}} \cr
& = & g_n^3 + 3g_n^2pa_n + 3g_n^2pb_n + 3g_npa_n^2 + 3g_npb_n^2 + 6g_npa_npb_n + 3g_n^2r_n + pa_n^3 + pb_n^3 + 27pc_n^3 \cr
& & + pd_n^3 + 3pa_n^2pb_n + 3pb_n^2pa_n + 27pc_n^2pd_n + 9pd_n^2pc_n + 6g_npa_nr_n + 6g_npb_nr_n + 3pa_n^2r_n \cr
& & + 3pb_n^2r_n + 6pa_npb_nr_n + 3g_nr_n^2 + 3pa_nr_n^2 + 3pb_nr_n^2 + r_n^3
\ ,
\label{pcpdeveq}
\eeqs
\beqs
r_{n+1} & = & g_npa_n^2 + g_npb_n^2 + 2g_npa_npb_n + pa_n^3 + pb_n^3 + 3pa_n^2pb_n + 3pa_n^2pc_n + pa_n^2pd_n + 3pb_n^2pa_n \cr
& & + 3pb_n^2pc_n + pb_n^2pd_n + 6pa_npb_npc_n + 2pa_npb_npd_n + 4g_npa_nr_n + 4g_npb_nr_n + 5pa_n^2r_n \cr
& & + 5pb_n^2r_n + 10pa_npb_nr_n + 12pa_npc_nr_n + 4pa_npd_nr_n + 12pb_npc_nr_n + 4pb_npd_nr_n \cr
& & + 4g_nr_n^2 + 8pa_nr_n^2 + 8pb_nr_n^2 + 12pc_nr_n^2 + 4pd_nr_n^2 + 4r_n^3
\ .
\label{reveq}
\eeqs

\section{Recursion relations for the ice model on $SG_3(n)$}

We give the recursion relations for the ice model on the generalized two-dimensional Sierpinski gasket $SG_3(n)$ here. Since the subscript is $b=3$ for all the quantities throughout this section, we will use the simplified notation $pa_{n+1}$ to denote $pa_3(n+1)$ and similar notations for other quantities. For any non-negative integer $n$, we have
\beqs
pa_{n+1}
& = & 3pa_n^6 + pb_n^6 + 16pa_n^5pb_n + 6pa_n^5pc_n + 2pa_n^5pd_n + 8pb_n^5pa_n + 35pa_n^4pb_n^2 + 54pa_n^4pc_n^2 \cr
& & + 6pa_n^4pd_n^2 + 25pb_n^4pa_n^2 + 54pb_n^4pc_n^2 + 6pb_n^4pd_n^2 + 40pa_n^3pb_n^3 + 108pa_n^3pc_n^3 + 4pa_n^3pd_n^3 \cr
& & + 108pb_n^3pc_n^3 + 4pb_n^3pd_n^3 + 24pa_n^4pb_npc_n + 8pa_n^4pb_npd_n + 36pa_n^4pc_npd_n + 6pb_n^4pa_npc_n \cr
& & + 2pb_n^4pa_npd_n + 36pb_n^4pc_npd_n + 36pa_n^3pb_n^2pc_n + 12pa_n^3pb_n^2pd_n + 216pa_n^3pc_n^2pb_n \cr
& & + 108pa_n^3pc_n^2pd_n + 24pa_n^3pd_n^2pb_n + 36pa_n^3pd_n^2pc_n + 24pb_n^3pa_n^2pc_n + 8pb_n^3pa_n^2pd_n \cr
& & + 216pb_n^3pc_n^2pa_n + 108pb_n^3pc_n^2pd_n + 24pb_n^3pd_n^2pa_n + 36pb_n^3pd_n^2pc_n + 324pc_n^3pa_n^2pb_n \cr
& & + 324pc_n^3pb_n^2pa_n + 12pd_n^3pa_n^2pb_n + 12pd_n^3pb_n^2pa_n + 324pa_n^2pb_n^2pc_n^2 + 36pa_n^2pb_n^2pd_n^2 \cr
& & + 144pa_n^3pb_npc_npd_n + 144pb_n^3pa_npc_npd_n + 216pa_n^2pb_n^2pc_npd_n + 324pa_n^2pc_n^2pb_npd_n \cr
& & + 108pa_n^2pd_n^2pb_npc_n + 324pb_n^2pc_n^2pa_npd_n + 108pb_n^2pd_n^2pa_npc_n \ ,
\label{pa3eq}
\eeqs
\beqs
pb_{n+1}
& = & pa_n^6 + 3pb_n^6 + 8pa_n^5pb_n + 16pb_n^5pa_n + 6pb_n^5pc_n + 2pb_n^5pd_n + 25pa_n^4pb_n^2 + 54pa_n^4pc_n^2 \cr
& & + 6pa_n^4pd_n^2 + 35pb_n^4pa_n^2 + 54pb_n^4pc_n^2 + 6pb_n^4pd_n^2 + 40pa_n^3pb_n^3 + 108pa_n^3pc_n^3 + 4pa_n^3pd_n^3 \cr
& & + 108pb_n^3pc_n^3 + 4pb_n^3pd_n^3 + 6pa_n^4pb_npc_n + 2pa_n^4pb_npd_n + 36pa_n^4pc_npd_n + 24pb_n^4pa_npc_n \cr
& & + 8pb_n^4pa_npd_n + 36pb_n^4pc_npd_n + 24pa_n^3pb_n^2pc_n + 8pa_n^3pb_n^2pd_n + 216pa_n^3pc_n^2pb_n \cr
& & + 108pa_n^3pc_n^2pd_n + 24pa_n^3pd_n^2pb_n + 36pa_n^3pd_n^2pc_n + 36pb_n^3pa_n^2pc_n + 12pb_n^3pa_n^2pd_n \cr
& & + 216pb_n^3pc_n^2pa_n + 108pb_n^3pc_n^2pd_n + 24pb_n^3pd_n^2pa_n + 36pb_n^3pd_n^2pc_n + 324pc_n^3pa_n^2pb_n \cr
& & + 324pc_n^3pb_n^2pa_n + 12pd_n^3pa_n^2pb_n + 12pd_n^3pb_n^2pa_n + 324pa_n^2pb_n^2pc_n^2 + 36pa_n^2pb_n^2pd_n^2 \cr
& & + 144pa_n^3pb_npc_npd_n + 144pb_n^3pa_npc_npd_n + 216pa_n^2pb_n^2pc_npd_n + 324pa_n^2pc_n^2pb_npd_n \cr
& & + 108pa_n^2pd_n^2pb_npc_n + 324pb_n^2pc_n^2pa_npd_n + 108pb_n^2pd_n^2pa_npc_n \ , 
\label{pb3eq}
\eeqs
\beqs
pc_{n+1}
& = & 5832pc_n^6 + 8pd_n^6 + pa_n^5pb_n + 3pa_n^5pc_n + pa_n^5pd_n + pb_n^5pa_n + 3pb_n^5pc_n + pb_n^5pd_n \cr
& & + 11664pc_n^5pd_n + 144pd_n^5pc_n + 4pa_n^4pb_n^2 + 4pb_n^4pa_n^2 + 9720pc_n^4pd_n^2 + 1080pd_n^4pc_n^2 \cr
& & + 6pa_n^3pb_n^3 + 162pa_n^3pc_n^3 + 6pa_n^3pd_n^3 + 162pb_n^3pc_n^3 + 6pb_n^3pd_n^3 + 4320pc_n^3pd_n^3 \cr
& & + 21pa_n^4pb_npc_n + 7pa_n^4pb_npd_n + 21pb_n^4pa_npc_n + 7pb_n^4pa_npd_n + 48pa_n^3pb_n^2pc_n \cr
& & + 16pa_n^3pb_n^2pd_n  + 162pa_n^3pc_n^2pd_n + 54pa_n^3pd_n^2pc_n + 48pb_n^3pa_n^2pc_n + 16pb_n^3pa_n^2pd_n \cr
& & + 162pb_n^3pc_n^2pd_n + 54pb_n^3pd_n^2pc_n + 486pc_n^3pa_n^2pb_n + 486pc_n^3pb_n^2pa_n + 18pd_n^3pa_n^2pb_n \cr
& & + 18pd_n^3pb_n^2pa_n + 486pa_n^2pc_n^2pb_npd_n + 162pa_n^2pd_n^2pb_npc_n + 486pb_n^2pc_n^2pa_npd_n \cr
& & + 162pb_n^2pd_n^2pa_npc_n \ ,
\label{pc3eq}
\eeqs
\beqs
pd_{n+1}
& = & pa_n^6 + pb_n^6 + 5832pc_n^6 + 8pd_n^6 + 3pa_n^5pb_n + 9pa_n^5pc_n + 3pa_n^5pd_n + 3pb_n^5pa_n + 9pb_n^5pc_n \cr
& & + 3pb_n^5pd_n + 11664pc_n^5pd_n + 144pd_n^5pc_n + 3pa_n^4pb_n^2 + 3pb_n^4pa_n^2 + 9720pc_n^4pd_n^2 \cr
& & + 1080pd_n^4pc_n^2 + 2pa_n^3pb_n^3 + 162pa_n^3pc_n^3 + 6pa_n^3pd_n^3 + 162pb_n^3pc_n^3 + 6pb_n^3pd_n^3 \cr
& & + 4320pc_n^3pd_n^3 + 27pa_n^4pb_npc_n + 9pa_n^4pb_npd_n + 27pb_n^4pa_npc_n + 9pb_n^4pa_npd_n \cr
& & + 36pa_n^3pb_n^2pc_n + 12pa_n^3pb_n^2pd_n + 162pa_n^3pc_n^2pd_n + 54pa_n^3pd_n^2pc_n + 36pb_n^3pa_n^2pc_n \cr
& & + 12pb_n^3pa_n^2pd_n + 162pb_n^3pc_n^2pd_n + 54pb_n^3pd_n^2pc_n + 486pc_n^3pa_n^2pb_n + 486pc_n^3pb_n^2pa_n \cr
& & + 18pd_n^3pa_n^2pb_n + 18pd_n^3pb_n^2pa_n  + 486pa_n^2pc_n^2pb_npd_n + 162pa_n^2pd_n^2pb_npc_n \cr
& & + 486pb_n^2pc_n^2pa_npd_n + 162pb_n^2pd_n^2pa_npc_n \ . 
\label{t23eq}
\eeqs

\end{document}